\documentclass[%
 amsmath,amssymb,
 aps,
pra,
]{revtex4}

\usepackage{graphicx}
\usepackage[caption=false]{subfig}
\usepackage{float}   
\usepackage{xcolor}

\begin{document}


\title{Three-body and Coulomb interactions in a quasi-two-dimensional dipolar Bose condensed gas}

\author{S. Mostafa Moniri}%
\email{s.m.moniri@iut.ac.ir}
\affiliation{Basic Sciences Group, Golpayegan College of Engineering, Isfahan University of Technology, Golpayegan 87717-67498, Iran}
\author{Heshmatollah Yavari}%
\email{h.yavary@sci.ui.ac.ir}
\affiliation{Department of Physics, University of Isfahan, Isfahan 81746, Iran}
\author{Elnaz Darsheshdar}
\email{darsheshdare@gmail.com}
\affiliation{%
Departamento de F\'{i}sica, Universidade Federal de S\~{a}o Carlos,
P.O. Box 676, 13565-905, S\~{a}o Carlos, S\~{a}o Paulo, Brazil
}%




\date{\today}

\begin{abstract}
In this paper, we studied a dilute quasi-two-dimensional dipolar Bose-condensed with two- and three-body contact, and Coulomb interactions using the Hartree-Fock-Bogoliubov-Popov approximation. We analyze numerically the effects of three-body contact, and Coulomb interactions on the energy spectrum, the quantum and thermal noncondensate fraction of the system. We show that increasing the three-body contact and Coulomb interactions leads to the appearance of rotonization and condensate instability at stronger dipole-dipole interaction. Also we find that the temperature dependence of the thermal noncondensate fraction is linear at low temperature. 
\begin{description}
\item[PACS numbers]
03.75.Nt, 05.30.Jp.
\end{description}
\end{abstract}

\maketitle


\section{Introduction}
Since the first observation of the ultracold dipolar Bose gases, the effect of long range interaction which induces strong correlations in these systems has attracted great attentions in theoretical and experimental studies \cite{Baranov2008,Lahaye2009,Carr2009,Baranov2012}. Based on the previous researches the s-wave scattering approximation neglects microscopic details of the two-particle interaction and replaces the true interaction with a contact pseudopotential interaction. To describe the effects of microscopic structure of the two body potential on the properties of these systems, it is necessary to go beyond the s-wave approximation. 

Ultracold gases of dipolar particles, that are interacting via long range anistropic dipole-dipole potential, remarkably changed the nature of the quantum degenerate regimes and opened interesting perspective for the observations of new physical phenomena \cite{Kadau2016}.
More specifically the two-dimensional (2D) dipolar Bose system with bosonic dipoles perpendicular to the plane of their translatory motion can undergo a phase transition to the Bose-Einstein condensation (BEC) and also obtain a roton-maxon excitation spectrum which is well known in the physics of the liquid helium due to momentum dependence of dipole-dipole interaction (DDI) \cite{Kapitza1938,Landau1941,Feynman1957}.

On the other hand, in addition to the two body interactions, the three-body interactions (TBI) have also an important role in a wide variety of interesting physical phenomena and provide a new physics compared to the two-body interactions. The  TBI realized in ultracold Bose gas by both experimental and theoretical methods \cite{Hammer2013,Will2010,Daley2014,Petrov2014}. 
Interplay between attractive two-body and large repulsive TBI may lead to new phenomena in weakly interacting Bose and Fermi gases \cite{Bulgac2002}. 
The effects of the TBI on the collective exitations \cite{Abdullaev2001,Jibbouri2013,Li2010}, the transition temperature, the condensate depletion, the stability of Bose condensate \cite{Ping2009,Mashayekhi2013} and the ultracold bosonic atoms loaded in an optical lattice \cite{Daley2009,Mazza2010,Singh2012,Mahmud2013} have been investigated so far.  
 
Observation of a remarkable quantum phenomenon that combines superfluidity with a crystalline order i.e., supersolidity in ultracold dipolar gases opens a new windows in both theoretical and experimental researches \cite{Kuklov2011,Boninsegni2012}. Since the collapsing gas prevents to achieve supersolid phase in dilute 2D dipolar Bose gases it is still an open question whether a supersolids can exist in the dilute regime or not. A stable supersolid state in a dilute 2D dipolar system was predicted by including a contact repulsive TBI, which may prevent the collapse \cite{Lu2015}.

The effects of TBI on a trapped dipolar Bose gas at finite temperature using the Hartree-Fock-Bogoliubov (HFB) approximation was investigated \cite{Boudjemaa2017}. The impacts of weak disorder with Gaussian correlation function on a dipolar Bose gas with TBI using the HFB theory have also studied \cite{Keltoum2019}. Combined effects of the TBI and DDI on the condition of modulational instability of an optically-trapped dipolar BEC in the mean field level was analytically studied as well \cite{Qi2018}. 
 
Considering the above interesting phenomena in the presence of TBI, studies on the effect of this interaction on the properties of the systems still have broad potential in ultracold atoms.  
On the other hand, another long-range interaction is described by the pair potential which is proportional to $C/r$. It is a repulsive Coulomb interaction of charged atoms ($C > 0$) and, in addition to the TBI, is the focus of this paper. The effect of this interaction has been studied on the transport properties \cite{Darsheshdar2016},  damping \cite{Moniri2016} and analytical depletion of the condensates \cite{Tamaddonpur2019} as well as the appearance of the bipolarons \cite{Hague2007} and the ground state energy of the system \cite{Lieb2001}. To the best of our knowledge the effect of Coulomb interaction in the presence of TBI on the depletion and energy spectrum or the roton-maxon structure of the dipolar Bose condensate have never been studied.
 
In this paper we study the effects of the three-body contact and long-rang Coulomb interactions interactions on the superfluid properties of a 2D dipolar Bose condensed gas  in the framework of HFB-Popov approximation.

\section{Formulation of the Problem}
We consider a dilute Bose-condensed gas of dipolar bosons with two- and three-body contact, and Coulomb interactions. The condensate is confined in one direction (z) and the dipole moments $d$ are oriented perpendicular to the x-y plane. The second quantized Hamiltonian of this 2D system is
\begin{align} \label{eq:1} 
&\mathcal{H} = \int d^2 r \{ \Psi^{\dagger}(\mathbf{r}) \left[ -\frac{\hbar^2\nabla^2}{2m} - \mu + U(\textbf{r}) \right] \Psi(\mathbf{r}) + \frac{1}{2} \int d^2 r' V_{2D}(\mathbf{r}-\mathbf{r'}) \Psi^{\dagger}(\mathbf{r}) \Psi^{\dagger}(\mathbf{r'}) \Psi(\mathbf{r'}) \Psi(\mathbf{r}) \nonumber \\
& + \frac{g_3}{6} \Psi^{\dagger}(\mathbf{r}) \Psi^{\dagger}(\mathbf{r})\Psi^{\dagger}(\mathbf{r}) \Psi(\mathbf{r}) \Psi(\mathbf{r}) \Psi(\mathbf{r}) \},
\end{align}
where $\Psi^{\dagger}(\mathbf{r})$ and $\Psi(\mathbf{r}) $ are respectively the usual creation and annihilation field operators, $m$ is the particle mass, $\mu $ is the chemical potential and ${{g}_{3}}$ is the strength of  contact repulsive TBI.  The 2D interaction potential ${{V}_{2D}}(\mathbf{r})$ includes contact, dipole-dipole and Coulomb interactions,
\begin{align} \label{eq:2} 
{{V}_{2D}}(\mathbf{r}) = g_{2D} \delta (\mathbf{r}) + U_{2D}^{dd}(\mathbf{r}) + U_{2D}^{C}(\mathbf{r}),
\end{align}
where $g_{2D}$ is the 2D short-range coupling constant, $U_{2D}^{dd}(\mathbf{r})$ is 2D-DDI and $U_{2D}^{C}(\mathbf{r})$ is the Coloumb interaction. In general the three-body coupling constant interaction ${{g}_{3}}$ is a complex number where its imaginary part  describing the three-body recombination loss and its real part content the three-body scattering parameter. 
We use the usual treatment for Bose systems with broken gauge symmetry and decomposed the bosonic operator as a sum of two parts,
\begin{align}  \label{eq:3} 
\Psi (\mathbf{r})=\phi (\mathbf{r})+\psi (\mathbf{r}),
\end{align}
where $  \phi (\mathbf{r})= \langle \Psi (\mathbf{r})  \rangle  $ is the condensate wave function, and the operator  $  \psi (\mathbf{r}) $ acts on the noncondensed particles which by definition have the property of $ \langle \psi (\mathbf{r})  \rangle =0 $. By applying the decomposition (\ref{eq:3}) in Eq. (\ref{eq:1}) and expanding the resulted expression and considering  only the terms quadratic and quartic in the noncondensate operators and ignoring  all averages of cubic products of the noncondensate operators, corresponding Hamiltonian for the noncondensed atoms can be written as,
\begin{align}  \label{eq:4} 
&\mathcal{H}_{nc} = \int{d^2r}{{\psi }^{\dagger }}(\mathbf{r})\left[ -\frac{\hbar^2{\nabla }^{2}}{2m}-\mu + U(\textbf{r}) \right]\psi (\mathbf{r})+ \nonumber \\ 
& +\frac{1}{2}\int {d^2r}d^2{r}'{{V}_{2D}}((\mathbf{r}-(\mathbf{r'})  [\phi(\mathbf{r}) \phi (\mathbf{r'}) \psi^{\dagger}(\mathbf{r}) \psi^{\dagger}(\mathbf{r'}) + \phi^*(\mathbf{r}) \phi^*(\mathbf{r'}) \psi(\mathbf{r'}) \psi (\mathbf{r}) \nonumber \\ 
&  +\phi^*(\mathbf{r}) \phi(\mathbf{r'}) \psi^{\dagger}(\mathbf{r'}) \psi (\mathbf{r}) + \phi(\mathbf{r}) \phi^*(\mathbf{r'}) \psi^{\dagger}(\mathbf{r}) \psi(\mathbf{r'}) +\phi(\mathbf{r'}) \phi^*(\mathbf{r'}) \psi^{\dagger}(\mathbf{r}) \psi (\mathbf{r}) \nonumber \\ 
& + \phi(\mathbf{r}) \phi^*(\mathbf{r}) \psi^{\dagger}(\mathbf{r'}) \psi(\mathbf{r'}) +\psi^{\dagger}(\mathbf{r}) \psi^{\dagger}(\mathbf{r'}) \psi(\mathbf{r'}) \psi (\mathbf{r})]  \nonumber \\ 
& +\frac{g_3}{6} \int {d^2r} [3\phi^2(\mathbf{r}) \psi^{\dagger}(\mathbf{r})\psi^{\dagger}(\mathbf{r})\psi^{\dagger}(\mathbf{r}) \psi(\mathbf{r}) + 3\phi^{*2} (\mathbf{r}) \psi^{\dagger}(\mathbf{r}) \psi(\mathbf{r})\psi(\mathbf{r})\psi(\mathbf{r}) \nonumber \\ 
& + 3\phi(\mathbf{r}) \phi^{*3}(\mathbf{r}) \psi(\mathbf{r})\psi(\mathbf{r}) + 3\phi^3(\mathbf{r}) \phi^*(\mathbf{r}) \psi^{\dagger}(\mathbf{r})\psi^{\dagger}(\mathbf{r})  + 9\phi(\mathbf{r}) \phi^*(\mathbf{r}) \psi^{\dagger}(\mathbf{r}) \psi^{\dagger}(\mathbf{r}) \psi(\mathbf{r}) \psi(\mathbf{r}) \nonumber \\ 
&+ 9\phi^2(\mathbf{r}) \phi^{*2}(\mathbf{r}) \psi^{\dagger}(\mathbf{r}) \psi (\mathbf{r}) + \psi^{\dagger}(\mathbf{r})\psi^{\dagger}(\mathbf{r})\psi^{\dagger}(\mathbf{r})\psi (\mathbf{r})\psi (\mathbf{r})\psi (\mathbf{r})].
\end{align}
For simplicity the condensed, normal and anomalous noncondensed densities are defined as,
\begin{align}  \label{eq:5} 
& n_c(\mathbf{r})= \langle {{\phi }^{*}}(\mathbf{r})\phi (\mathbf{r})  \rangle, \nonumber\\
& n(\mathbf{r})= \langle {{\psi }^{\dagger }}(\mathbf{r})\psi (\mathbf{r})  \rangle, \nonumber\\
& n(\mathbf{{r}'},\mathbf{r})= \langle {{\psi }^{\dagger }}(\mathbf{{r}'})\psi (\mathbf{r})  \rangle, \nonumber\\
& m(\mathbf{{r}'},\mathbf{r})= \langle \psi (\mathbf{{r}'})\psi (\mathbf{r})  \rangle. 
\end{align}
In the mean-field approximation the quartic and higher order product of the noncondensate operators may be written as,
\begin{align}\label{eq:6} 
& {{\psi }^{\dagger }}{{{{\psi }'}}^{\dagger }}{\psi }'\psi =\left\langle {{{{\psi }'}}^{\dagger }}{\psi }' \right\rangle {{\psi }^{\dagger }}\psi +\left\langle {{\psi }^{\dagger }}{\psi }' \right\rangle {{{{\psi }'}}^{\dagger }}\psi +\left\langle {{{{\psi }'}}^{\dagger }}\psi  \right\rangle {{\psi }^{\dagger }}{\psi }'+\left\langle {{\psi }^{\dagger }}\psi  \right\rangle {{{{\psi }'}}^{\dagger }}{\psi }', \nonumber\\ 
& {{\psi }^{\dagger }}{{\psi }^{\dagger }}\psi \psi =4\left\langle {{\psi }^{\dagger }}\psi  \right\rangle {{\psi }^{\dagger }}\psi \quad,\quad {{\psi }^{\dagger }}{{\psi }^{\dagger }}{{\psi }^{\dagger }}\psi \psi \psi =36{{\left\langle {{\psi }^{\dagger }}\psi  \right\rangle }^{2}}{{\psi }^{\dagger }}\psi,  \nonumber\\ 
& {{\psi }^{\dagger }}{{\psi }}{{\psi }}{{\psi }}=3\left\langle {{\psi }^{\dagger }}\psi  \right\rangle \psi \psi  \quad,\quad {{\psi }^{\dagger}}{{\psi }^{\dagger}}{{\psi }^{\dagger}}\psi =3\left\langle {{\psi }^{\dagger }}\psi  \right\rangle {{\psi }^{\dagger }}{{\psi }^{\dagger }}. 
\end{align}
 To obtain the mean-field factorizations (\ref{eq:6}) we have neglected the terms proportional to the anomalous non-condensate density $m(\mathbf{{r}'},\mathbf{r})=\left\langle \psi (\mathbf{{r}'})\psi(\mathbf{r}) \right\rangle$ and to its complex conjugate. This approximation corresponds to the so called Popov approximation \cite{Popov1972} and is expected to be appropriate both at high temperatures, where $n(r)\gg m(r)$, and low temperatures where the two densities are of the same order, but both negligibly small for the very dilute systems \cite{Giorgini1997}. Here we assume $\phi (\mathbf{r})$  to be real without any loss of generality \cite{Natu2013}. Then, the Hamiltonian (\ref{eq:4}) can be written as,
\begin{align}\label{eq:7} 
& \mathcal{H}_{nc}=\int{{{d}^{2}}r}{{\psi }^{\dagger }}(\mathbf{r}){{{\hat{L}}}_{0}}\,\psi (\mathbf{r}) +\frac{1}{2}\int{{{d}^{2}}r}{{d}^{2}}{r}'{{V}_{2D}}(\mathbf{r}-\mathbf{{r}'})\{{{n}_{tot}}(\mathbf{r},\mathbf{{r}'})[{{\psi }^{\dagger }}\left( {\mathbf{{r}'}} \right)\psi (\mathbf{r})+{{\psi }^{\dagger }}\left( \mathbf{r} \right)\psi (\mathbf{{r}'})] \nonumber \\  
& +\phi (\mathbf{r})\phi (\mathbf{{r}'})[{{\psi }^{\dagger }}(\mathbf{r}){{\psi }^{\dagger }}(\mathbf{{r}'})+\psi (\mathbf{{r}'})\psi (\mathbf{r})]\} + \int{{{d}^{2}}r} \left[ 9{{n}_{c}}(\mathbf{r})n(\mathbf{r})+3n_{c}^{2}(\mathbf{r}) \right]\left[ {{\psi }^{\dagger }}(\mathbf{r}){{\psi }^{\dagger }}(\mathbf{{r}})+\psi (\mathbf{{r}})\psi (\mathbf{r}) \right],
\end{align}
where ${{n}_{tot}}\left( \mathbf{r},\mathbf{{r}'} \right)={{n}_{c}}\left( \mathbf{r},\mathbf{{r}'} \right)+n\left( \mathbf{r},\mathbf{{r}'} \right)$ is the total density and 
\begin{align}\label{eq:8} 
 {{{\hat{L}}}_{0}}=-\frac{{\hbar^2{\nabla }^{2}}}{2m}-\mu + U(\mathbf{r})+\int{d\mathbf{{r}'}}{{V}_{2D}}(\mathbf{r}-\mathbf{{r}'}){{n}_{tot}}(\mathbf{{r}'})+\frac{{{g}_{3}}}{6}\left[ 9n_{c}^{2}(\mathbf{r})+36{{n}_{c}}(\mathbf{r}){n}(\mathbf{r}) + 36n^{2}(\mathbf{r}) \right]. 
\end{align}
In the Fourier space, the field operator of non-condensed atoms can be expanded in terms of plane waves $\psi (\mathbf{r})=(1/\sqrt{S}) \sum_k a_k {e}^{i\textbf{k.r}} $ where $S$ is the surface area and  $a_k$  is the annihilation particle operator. {We consider our quasi-2D system in the ultracold limit where the particle momentum satisfies the inequality $k r^* \ll 1$ with $r^*=m d^2/\hbar^2$ is the characteristic dipole-dipole distance.} The Fourier transform of DDI ($U^{dd}_{2D}({k})$) and Coulomb interaction ($U^{C}_{2D}({k})$) can be written as \cite{Boudjemaa2013},
\begin{align}\label{eq:15}
&U_{2D}^{dd}( {k} )=-2\pi {{d}^{2}}\left| {k} \right| = 2\pi \frac{\hbar^2 r^*}{m} \left| {k} \right|, \\ 
&U_{2D}^{C}( {k} )=2\pi \frac{C}{k}.
\end{align}
with $m$ the particle mass and $C$ the coupling constant. By assuming a weakly interacting system, where $mg_{2D}/2\pi \hbar^2 \ll 1$ and $r^* \ll \xi$ (with $\xi=\hbar / \sqrt{m n_c g_{2D}}$ being the healing length), the elementary excitations of the system can be found by the usual Bogoliubov transformation which diagonalizes the Hamiltonioan. Thus we can write $a_k=u_kb_k - v_kb^\dagger_{-k}$ and $a^\dagger_k =u_kb^\dagger_k - v_kb_{-k}$, with $b_k$ and $b^\dagger_k$ operators of the elementary excitations. The Bogoliubov coefficients $u_k$ and $v_k$ are
\begin{align}\label{eq:22}
&u_k =\sqrt{\frac{\epsilon_k+V_{tot}(k)}{2E_{k}} + \frac{1}{2}}, \\ \nonumber
&v_k =\sqrt{\frac{\epsilon_k+V_{tot}(k)}{2E_{k}} - \frac{1}{2}},
\end{align}
where  $\epsilon_k=\hbar^2k^2/2m$ being the free particle energy, $E_k$ is the excitation energy, $C_d=2\pi {{d}^{2}}/g_{2D}$, $C_c=2\pi {C}/g_{2D}$ and we have $V_{tot}(k) =  n_c g_{2D} \left(1 - C_d k + C_c/k + g_3 n_c/g_{2D} \right) $.
For a uniform Bose condensate at low temperatures, where the condensate is not significantly depleted $E_k$ will be
\begin{align}\label{eq:16}
E_{k} = \bigg( \epsilon_k^2 + 2\epsilon_k \left[  n_c g_{2D}(1-C_dk + C_c/k) + g_3 n_c^2  \right]  \bigg)^{1/2}. 
\end{align}
For three dimensional case  $U_{3D}^{dd}\left( k \right)=\frac{{{C}_{dd}}}{3}\left( 3{{\cos }^{2}}{{\theta }_{k}}-1 \right)$ and ${{g}_{2D}}=g$ and in the absence of Coulomb interaction, Eq. (\ref{eq:16}) reduces to Eq. (10) of Ref. \cite{Boudjemaa2017A}. The diagonal form of the Hamiltonian (\ref{eq:1}) can be written as
\begin{align}
\mathcal{H} = E + \sum_{k} E_{k} b^\dagger_{k} b_{k},
\end{align}
where the total energy is $E =E_0+E'$. The zero order term $E_0$ is accounted for the condensate ($k \rightarrow 0$),
\begin{align}
E_0=\frac{1}{2} N_c n_c g_{2D}(1 + g_3 n_c/g_{2D}).
\end{align}
The ground-state energy correction due to quantum fluctuations $E'$ is
\begin{align}
E'=\frac{1}{2}\sum_k [E_k - \epsilon_k - n_c V_{tot}(k)].
\end{align}

In the low momentum $( k\to 0 )$, the dispersion relation (\ref{eq:16}) is a non-phononic kind of $E_k= \hbar \sqrt{n_c g_{2D}C_c k/m}$. On the other hand, in the absence of the Coulomb interaction the dispersion relation is similar to the sound wave i.e. $E_k= \hbar v_s k$ with sound velocity of
\begin{align}
v_s=\sqrt{\frac{{n_cg_{2D}}(1+{n_cg_3}/{g_{2D}})}{m}}.
\end{align}
It is convenient to represent Eq. (\ref{eq:16}) as a dimensionless form of
\begin{align}\label{eq:18}
 \frac{E_k}{n_cg_{2D}}  =\frac{1}{2} \xi k \left[\left(\xi k -\frac{2 C_d}{\xi }\right)^2 + 4 \left(1 + \frac{C_c}{k} + \frac{n_c g_3}{g_{2D}}\right) - \frac{4 C_d^2}{\xi ^2} \right]^{1/2}. 
\end{align}

\begin{figure}[H]
\centering
	\subfloat[]{
		\includegraphics[width=0.4\textwidth]{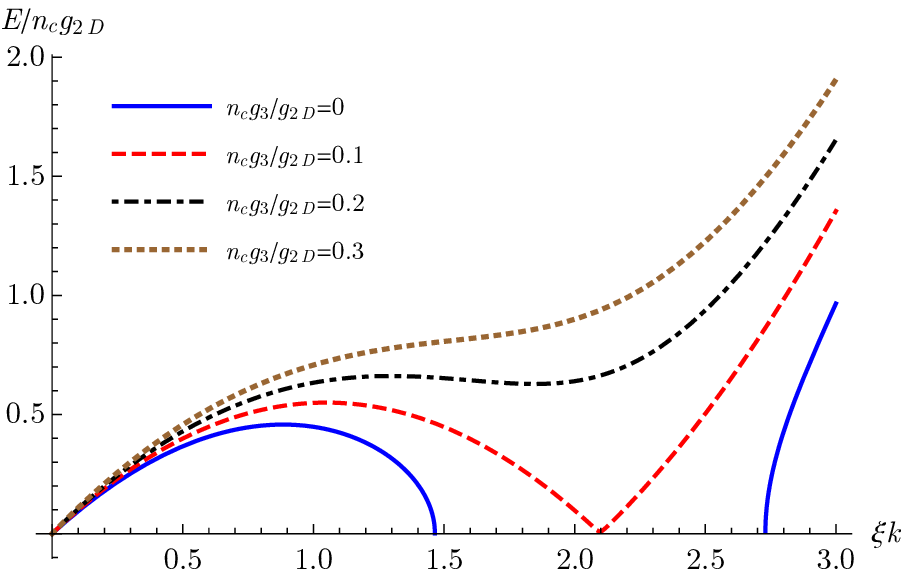}
        \label{fig:1a}}
	\subfloat[]{
		\includegraphics[width=0.4\textwidth]{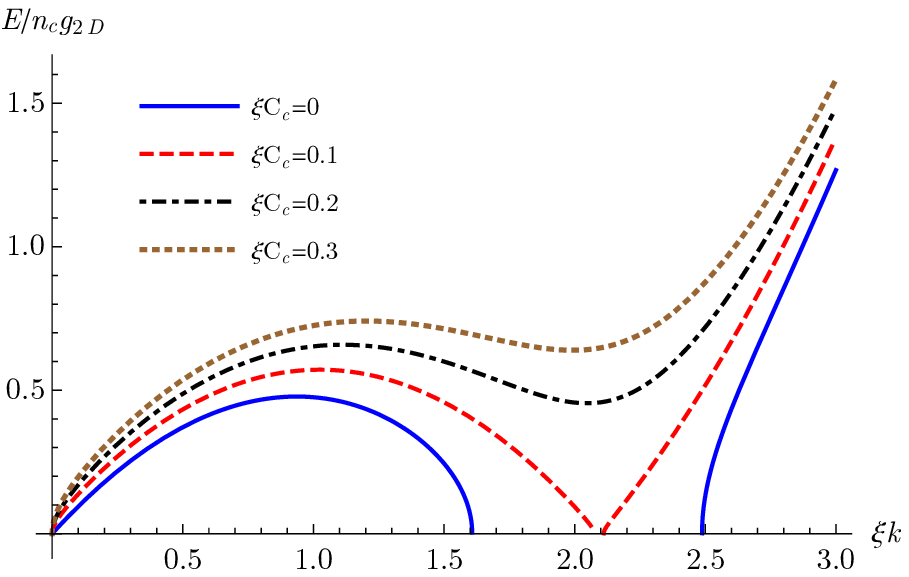}
        \label{fig:1b}}
\caption{Excitation energy $E_k$ of a quasi-2D dipolar BEC as a function of momentum $k$ for (a) different values of $n_cg_3/g_{2D}$   for the $C_d/\xi=1.049$ and  $\xi C_c=0$,  (b) different values of $\xi C_c$ for the $C_d/\xi = 1.023$ and $n_cg_3/g_{2D}=0$.}
\label{fig:1}
\end{figure}

In the absence of the Coulomb interaction, one can see from Eq. (\ref{eq:18}) that for $C_d<\frac{\sqrt{8}}{3}\xi(1+{n_cg_3}/{g_{2D}})^{1/2}$ the dispersion relation is a monotonic function of $k$ and it has a roton-maxon structure in the interval of $\frac{\sqrt{8}}{3}\xi(1+{n_cg_3}/{g_{2D}})^{1/2}< C_d < {\xi(1+{n_cg_3}/{g_{2D}})^{1/2}}$. In this case, in $C_d={\xi(1+{n_cg_3}/{g_{2D}})^{1/2}}$ the roton minimum is zero and at larger $C_d$ the Bose condensate is unstable. If the roton is close to $0$, then $k_r=2C_d/\xi^2$ is the position of the roton, and
\begin{align}
\Delta = \frac{\hbar^2}{2m}k_r\sqrt{\frac{4(1+{n_cg_3}/{g_{2D}})}{\xi^2}-k_r^2}
\end{align}
is the height of the roton minimum. In the presence of of Coulomb interaction, the value of $k_r$ and $\Delta$ can be obtaiend numerically. If $g_3=0$ and $C_c=0$, Eq. (\ref{eq:18}) reduces to Eq. (4) in \cite{Boudjemaa2016} and  Eq. (5) in \cite{Boudjemaa2013}.We plot the excitation energy  $E_k$ as a function of momentum $k$ for different values of $\xi C_c$ and $n_cg_3/g_{2D}$ in Fig. (\ref{fig:1}). It can be seen that both the three-body and Coulomb interactions modify the collective excitations and stabilization of the Bose condensate, this depends on their interaction strength. As it can be seen in Figs. (\ref{fig:1a}) and (\ref{fig:1b}), by increasing the strength of the three-body and Coulomb interactions the roton-maxon structure will be disappeared in  $C_d/\xi>\frac{\sqrt{8}}{3}$,  and the condensate instability happens at stronger dipole-dipole interaction. The TBI has stronger effect on the above issues compared to the Coulomb interaction.

The noncondensed density is defined as
\begin{align}
	n'= \sum_k \langle a_k^\dagger a_k \rangle = \frac{1}{S} \sum_k \left[ v_k^2 + \left( v_k^2 + u_k^2 \right) 		N_k \right],
\end{align}
where $N_k = \langle b_k^\dagger b_k \rangle = \left[ \textrm{Exp}(E_k/k_BT) -1 \right]^{-1}$ is the Bose-Einstein distribution function and $\langle b_k^\dagger b_k^\dagger \rangle = \langle b_k b_k \rangle = 0$.  Using the expressions of $u_k$ and $v_k$ and working in the thermodynamic limit where the sum over $k$ can be replaced by the integral, one can get $n' = n_0 + n_{th}$, where $n_0$ is the depletion due to the quantum fluctuations related to the contact interactions (two- and three-body), DDI and Coulomb interaction and $ n_{th}$ is the depletion due to thermal fluctuations. Then these depletions can be written as
\begin{align}\label{eq:25}
n_0=\frac{1}{2} \int \frac{d^2k}{(2\pi)^2} \left[ \frac{{{\epsilon}_{k}}+{{n}_{c}}{{g}_{2D}}\left( 1-C_dk+C_c/k+{{g}_{3}}{{n}_{c}}/{{g}_{2D}} \right)}{{{E}_{k}}}-1 \right], 
\end{align}
\begin{align}\label{eq:26}
n_{th} =  \int \frac{d^2k}{(2\pi)^2} \left[ \frac{{{\epsilon}_{k}}+{{n}_{c}}{{g}_{2D}}\left( 1-C_dk+C_c/k+{{g}_{3}}{{n}_{c}}/{{g}_{2D}} \right)}{{{E}_{k}}} \right]\frac{1}{{{e}^{{{E}_{k}}/{{k}_{B}}T}}-1}.  
\end{align}

\begin{figure}
\centering
	\subfloat[]{
		\includegraphics[width=0.4\textwidth]{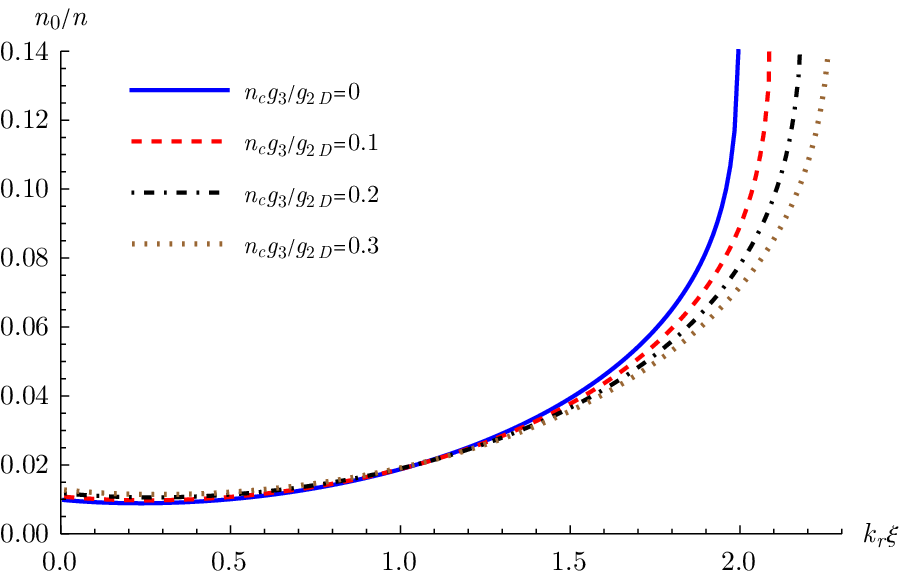}
        \label{fig:2a}}
	\subfloat[]{
		\includegraphics[width=0.4\textwidth]{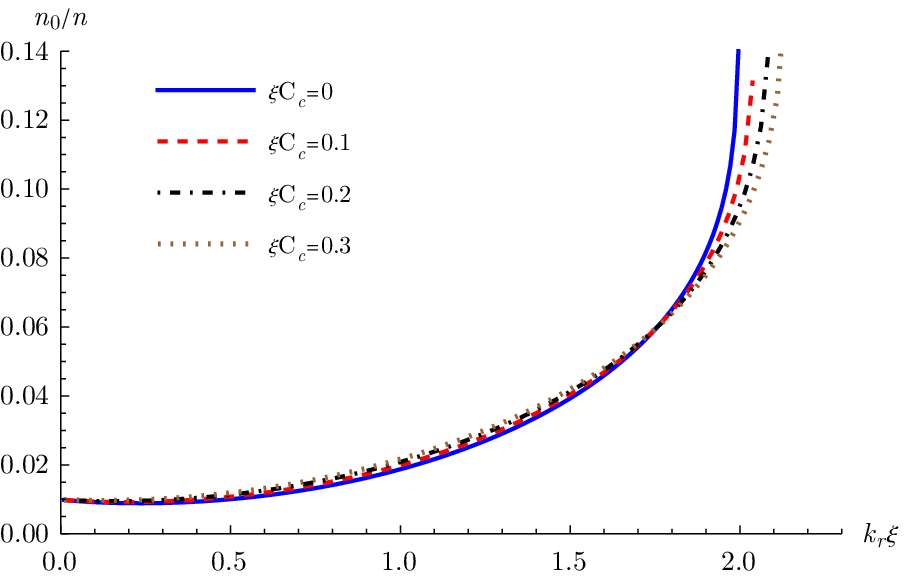}
        \label{fig:2b}}
\caption{ The quantum noncondensate fraction of a quasi-2D dipolar Bose gas at the presence of the contact (two- and three-body) and Coulomb interactions as a function of $k_r \xi$ for (a)  different value of $n_cg_3/g_{2D}$ with $\xi C_c=0$, and  (b) different value of $\xi C_c$ with $n_cg_3/g_{2D}=0$. Here we consider $1/r^*=100/(k_r\xi^2)$.}
\label{fig:2}
\end{figure}

The $C_d k$ term in the integral of Eq. (\ref{eq:25}) leads to a divergency at large momenta. To overcome this problem we set a cut-off equal to $1/r^*$. This is valid in the ultracold regime ($k \ll 1/r^*$). The quantum depletion of Eq. (\ref{eq:25}) is calculated numerically and the quantum noncondensate fraction is plotted as a function of $k_r \xi$ in the Fig. (\ref{fig:2}) for different value of $\xi C_c$ and $n_cg_3/g_{2D}$. In the absence of three-body and Coulomb interactions, the calculated quantum noncondensate fraction (solid blue line in Fig. (\ref{fig:2})) is in agreement with those obtained in \cite{Boudjemaa2013}. 
In addition both Figs. (\ref{fig:2}) and (\ref{fig:1}) confirm that the condensate instability happens at stronger dipole-diploe interaction for non-zero values of $\xi C_c$ and $n_cg_3/g_{2D}$. It can be seen from Fig. (\ref{fig:2a}) that at small values of $k_r \xi$ (weak DDI),  the TBI has more contribution in the depletion of the condensate and by increasing $k_r \xi$ up to $k_r \xi=1$, both TBI and DDI have the same contributions. At larger values of $k_r \xi$ (strong DDI), the DDI plays the main role in the depletion. The same holds true for the Coulomb interaction in Fig. (\ref{fig:2a}) where DDI and Coulomb interaction have the same contribution at $k_r \xi=1.7$.

\begin{figure}[H]
\centering
	\includegraphics[width=0.5\textwidth]{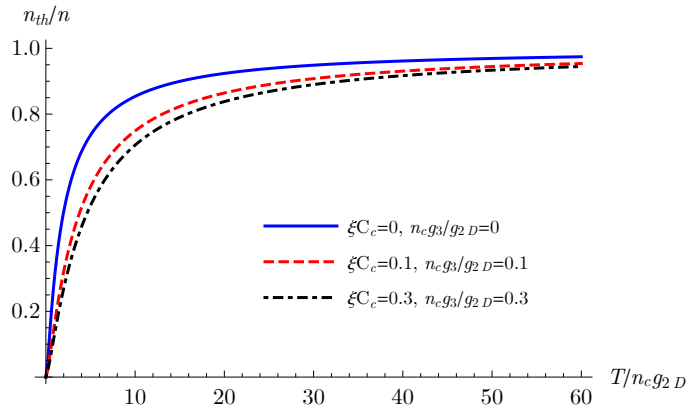}
    \caption{ The thermal noncondensate fraction of a quasi-2D dipolar Bose gas at the presence of the contact (two- and three-body) and Coulomb interactions as a function of the temperature for $k_r \xi=1.98$ and different values of $n_cg_3/g_{2D}$ and $\xi C_c$.}
	\label{fig:3}
\end{figure}

In the absence of the Coloumb interaction in the roton regime, $E_{k \approx k_r}= \Delta$, if the roton minimum is sufficiently close to zero ($\Delta \ll k_BT$) the  thermal noncondensate fraction can be obtained analytically as
\begin{align}\label{eq:27}
\frac{n_{th}}{n} = 	\frac{2}{\pi }\frac{mng_{2D}^{2}}{\hbar^2}\frac{k_BT}{\Delta^2}. 
\end{align}
{The obtained numerical results} for the thermal noncondensate fraction of Eq. (\ref{eq:26}) is shown in Fig. (\ref{fig:3}). As it can be seen {from Eq. (\ref{eq:27}) and} Fig. (\ref{fig:3}), the temperature dependence of the  thermal noncondensate fraction at low temperature is linear, which is in agreement with the analytical results in \cite{Tamaddonpur2019,Boudjemaa2013}. In addition, by increasing the three-body and Coulomb interactions,  decreasing the thermal noncondensate fraction is obvious. It should be noted that our calculations are in the HFB-Popov approximation that  is expected to be appropriate for both high and low temperatures. Then as is expected,  the thermal noncondensate fraction tends to unity at high temperature in Fig. (\ref{fig:3}). 

\section{Conclusion}
In the framework of HFB-Popov approximation, we analyzed the effects of the three-body and  Coulomb interactions on a trapped dipolar Bose gas at finite temperature. We showed that the intriguing interplay of the three body, coulomb and dipole-dipole interactions plays a fundamental role in the physics of the system. In the presence of the three-body and Coulomb interactions, the roton-maxon structure appears at the stronger dipole-dipole interaction. The same holds true for the condensate instability. In addition, the TBI has stronger effect compared to the Coulomb interaction in both of the above issues. The finite temperature noncondensate fraction shows a linear behavior at low temperature limits confirming an agreement with Refs. \cite{Tamaddonpur2019,Boudjemaa2013}. Increasing both the three-body and Coulomb interactions will decrease the finite temperature noncondensate fraction. 

\section*{Acknowledgement}
E. D. benefited from Grant from S\~ao Paulo Research Foundation (FAPESP, Grant Number 2018/10813-2).

\end{document}